\documentclass[%
 aip,
 jmp,%
 amsmath,amssymb,
 reprint,%
]{revtex4-1}

\usepackage{natbib,hyperref}
\hypersetup{
     colorlinks   = true,
     citecolor    = blue,
     urlcolor     = blue
}
\usepackage{graphicx}
\usepackage{dcolumn}
\usepackage{bm}
\newcommand{\IPCMS}{Institut de Physique et Chimie des Mat\'{e}riaux de Strasbourg, UMR 7504 CNRS, Universit\'{e} de Strasbourg, 23 rue du Loess, BP 43, 67034 Strasbourg Cedex 2, France}
\newcommand{\IJL}{Institut Jean Lamour, UMR 7198 Université de Lorraine, CNRS, Campus Artem, 2 allée André Guinier, BP 50840, 54011 Nancy Cedex, France}

\begin{document}

\preprint{AIP/123-QED}

\title{Increasing TeraHertz spintronic emission with planar antennas}

\author{Matthias~Pacé}
\affiliation{\IPCMS}
\author{Oleksandr~Kovalenko}
\altaffiliation{Currently at Faculty of Science and Technology, Lancaster University, United Kingdom}

\affiliation{\IPCMS}
\author{Jos\'{e}~Solano}
\affiliation{\IPCMS}
\author{Michel~Hehn}
\affiliation{\IJL}
\author{Matthieu~Bailleul}
\email{matthieu.bailleul@ipcms.unistra.fr}
\affiliation{\IPCMS}
\author{Mircea~Vomir}
\email{mircea.vomir@ipcms.unistra.fr}
\affiliation{\IPCMS}

\date{\today}

\begin{abstract}

Spintronic THz emitters, consisting of Ta/Co/Pt  trilayers patterned into rectangles of lateral size in the 10~µm range, have been integrated in planar electromagnetic antennas of various types (dipole, bow-tie, spiral). Antenna dimensions and  shapes have been optimized with the help of electromagnetic simulations so as to maximize antenna efficiency in both narrow-band and broad-band geometries at/around 1 THz. The THz emission has been studied using a pump probe free space electro-optic sampling set up, both for a single emitter geometry and for arrays of emitters. Results show an increase of the detected THz signal for all antenna geometries, with enhancement ratios in the range of three to fifteen depending on antenna type and frequency range, together with changes of the emission bandwidth consistent with simulated characteristics.

\end{abstract}

\maketitle


TeraHertz electromagnetic waves have numerous applications, be it for spectroscopy\cite{Jepsen2011,Mittleman2017} or next generation telecomunications\cite{Dang2020}. However, THz generation and detection remains a challenge, despite significant advances during the last years. The most convenient broad-band techniques are indirect time-domain methods relying on femtosecond laser pulses and their rectification in  non-linear crystals or photodiodes\cite{Schmuttenmaer2004}. An alternative method for THz emission\cite{Beaurepaire2004} involves the ultrafast demagnetization of magnetic materials\cite{Beaurepaire1996}. In engineered magnetic / non-magnetic metallic multilayers, this has been recently demonstrated to be very efficient.\cite{Seifert2016,Kampfrath2013} In this technique, a femtosecond laser pulse absorbed by a thin ferromagnetic layer generates a transient spin-polarized hot electron current\cite{Battiato2010,Melnikov2011,Choi2014,Huisman2016}, which upon injection in an adjacent non-magnetic layer is converted into an oscillating charge current via the spin orbit coupling, by virtue of the so-called inverse spin Hall effect\cite{Saitoh2006}. This ultrashort electrical current radiates in the free space and in the substrate in the form of a sub-picosecond electromagnetic pulse. Spintronic TeraHertz Emitters (STE) have a number of advantages: when using appropriate materials\cite{Bull2021} they can be extremely broadband and intense,\cite{Seifert2016,Seifert2022,Rouzegar2023} and it is relatively easy and cheap to grow the required thin films on a large variety of substrates, in contrast with electro-optical emitters which require either high quality bulk single crystals or specific photoconductive devices. In addition, the small volume of material involved in the conversion process suggests the possibility to miniaturize the technique down to the micrometer scale and possibly achieve on-chip THz spintronics.\cite{Jhuria2020,Hoppe2021} In this context, it is important to understand how to efficiently convert the local spintronic THz emission to free space radiation. Very recently, Nandi et al\cite{Nandi2019} and Talara et al.\cite{Talara2021} studied the change of emission using large-arms antennas directly adapted from those used for photoconductive techniques. In this study, we explore further the impact of the antenna geometry by comparing different designs and show that  far field THZ emission of STEs can be enhanced with antenna optimization.

\begin{figure}[h]
\includegraphics[width=8cm]{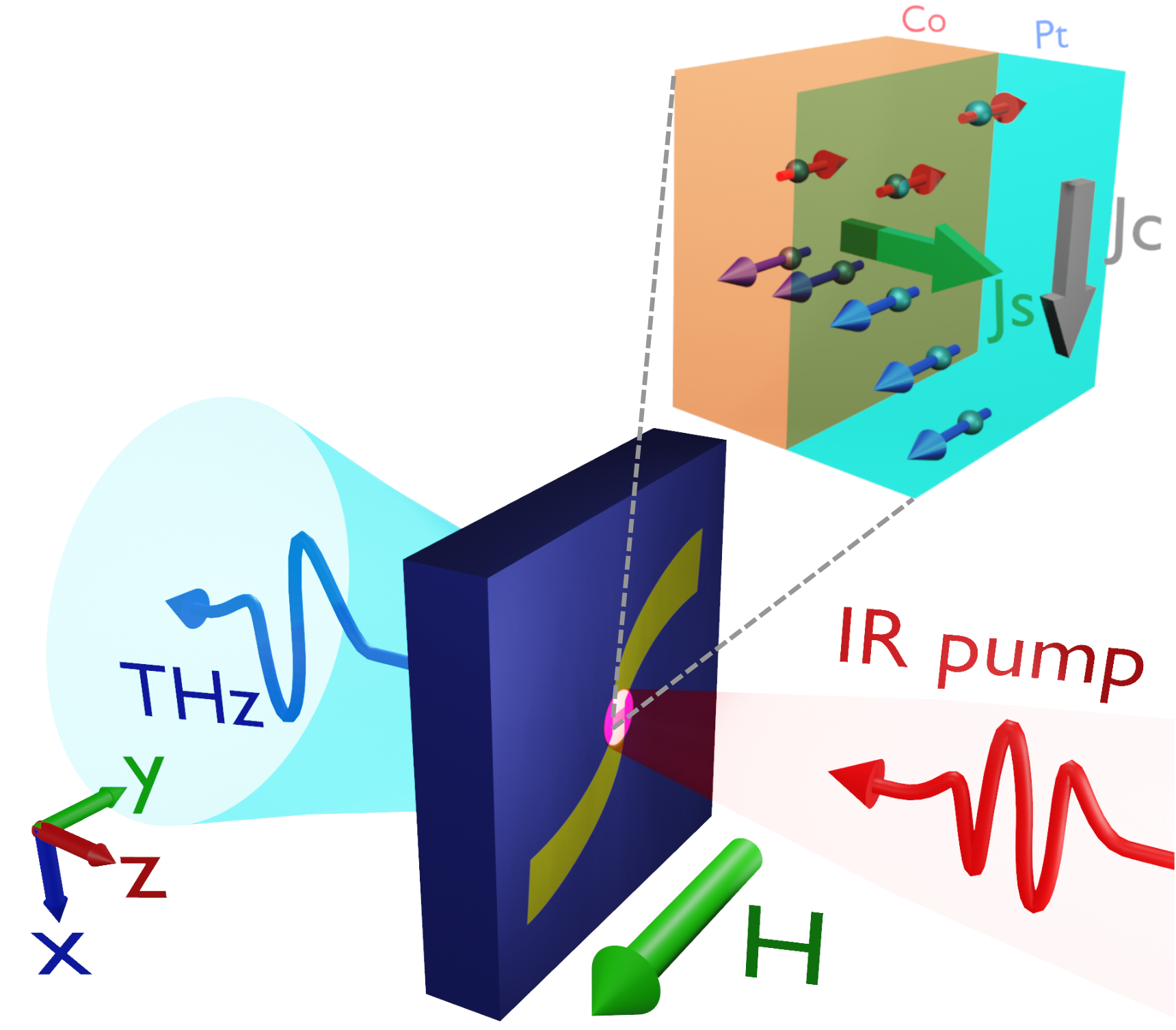}
\caption{(Color online) Sketch of the THz emission of a small area spintronic emitter equipped with an electromagnetic antenna. The inset illustrates the motion of spin-polarized electrons across the stack (the Ta underlayer is not represented here for simplicity).} \label{Fig1}
\end{figure}

Figure~\ref{Fig1} shows the sketch of the THz emission. The STE consists of a trilayer Ta(3)/Co(5)/Pt(3) (number in brakets are thickness in nm) deposited onto a sapphire substrate by magnetron sputtering and patterned into rectangles with sides in the range 7-15~µm. A planar antenna of appropriate shape and dimensions (see geometry in Fig. 2) is fabricated around it. The spintronic emitter is excited by femtosecond laser pulses  focused onto the substrate surface from the air side. The generated sub-ps pulses are collected from the substrate side and measured using electro-optic sampling (see In Fig. 1). 

\begin{figure}
    \centering
    \includegraphics[width=1\linewidth]{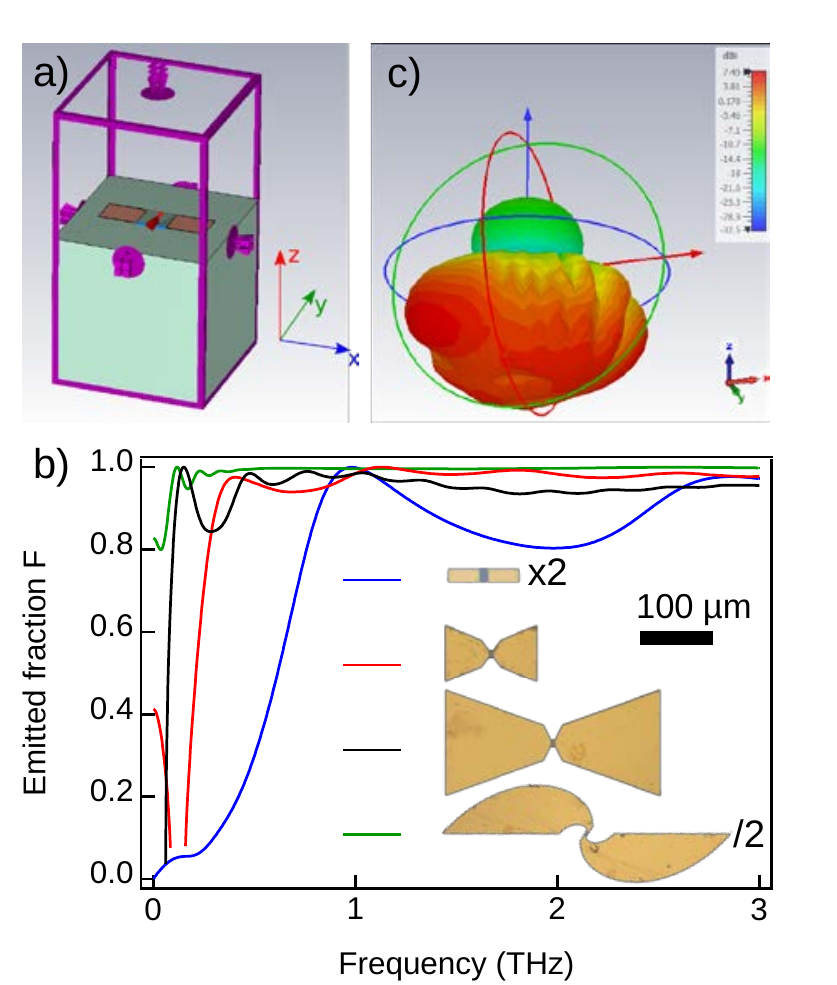}
    \caption{(a) Geometry of the electromagnetic simulation. (b) Simulated emission coefficients for the four designs shown in the legend (from top to bottom: dipole, compact bow-tie, large bow-tie, spiral, see scale bar and scale factors). (c) Radiation pattern for the spiral antenna at 2.5~THz.}
    \label{Fig2}
\end{figure}

Prior to fabrication, electromagnetic simulations have been conducted to optimize the antenna shape for emission at/around 1~THz using CST Studio Suite\cite{CSTstudio}. The STE is modelled as a local excitation port with an internal impedance $R_{\square}l/w$, where $R_{\square}=50$~$\Omega$, $l$ and $w$ are the sheet resistance, length and width of the trilayer stack, respectively. The substrate is assumed to be semi-infinite with a relative permittivity $\epsilon_r=9.7$, corresponding to sapphire, the antenna arms are idealized as perfect electric conductors and open boundary conditions are defined to emulate free space emission [see Fig.~\ref{Fig2}(a)]. The simulation proceeds by feeding a short voltage pulse (spectral content 0.1-3 THz) to this port and solving for the evolution of the electromagnetic field in the simulation box until the pulse is dissipated. The voltage reflected to the excitation port is calculated. Upon suitable Fourier transforms, the dependence of the complex reflection coefficient $S$ as function of frequency is determined. We calculate the average reflection coefficient over a frequency window centered around 1~THz and  look for its minimum as function of the geometrical parameters of the antenna. Material losses having been neglected, excepted in the excitation port itself, the fraction of signal which is emitted is $F=\sqrt{1-|S|^2}$. 

Fig.~\ref{Fig2}(b) shows the four optimized  geometries and the corresponding emission fraction as function of frequency. The first design is a simple dipole antenna with straight arms having the same width $w=10$ µm as the $l=7$µm long emitter. The optimization results in a total antenna length of 57 µm, very close to half the wavelength of an electromagnetic wave residing in both air and substrate  [Fig.~\ref{Fig2}(f)] ($\lambda/2=\frac{c}{f\sqrt{2(1+\epsilon_r)}}=64$ µm), thus bringing us to the half-wave antenna well-known in radio-frequency electronics.\cite{Kraus_book} As expected, the corresponding emission displays a main peak centered at the target frequency of 1 THz, and a relatively steep decrease on the low frequency side, with a 60 percents pass band starting at about 0.6 THz [blue line in Fig.~\ref{Fig2}(b)] . The other designs investigated are textbook geometries,\cite{Kraus_book} bow-tie\cite{Compton1987} and spiral.\cite{Dyson1959} Both involve to some extent ingredients ensuring broadband emission, namely scale-invariance (no preferred wavelength over a certain range) and self-complementarity (the metal area around the port having roughly the same shape as the no-metal area).\cite{Kraus_book} For the bow-tie design, starting from the same emitter size, we allowed for a two steps tapering, with the intermediate tapering line fixed at 10µm from the edge of the emitter. The optimization provided us with two solutions, each of them corresponding to a local minimum, namely a compact bow-tie (total length 127 µm) and a large bow-tie (total length 314 µm). Both designs provide a relatively featureless emission from 3 THz down to a  minimum frequency governed by the antenna dimension (60 percents pass band starting at about 0.2 and 0.08 THz for compact and large bow-ties, red and black lines in Fig.~\ref{Fig2}, respectively). The last design investigated is a so-called logarithmic spiral. The optimization returned a big antenna, spiralling over about 130° and with a maximum extension of about 920 µm, together with an elongated emitter with $w=7$ µm and $l=15$µm. The simulated THz emission has a flat spectral density and exceeds 90 percents above 0.1 THz (green line in Fig.~\ref{Fig2}).

For fabrication, we start with an extended trilayer Ta 3/Co 5/Pt 3 nm deposited onto a sapphire substrate by magnetron sputtering. The top Pt layer is chosen for its large spin Hall angle.\cite{Sinova2015} The bottom Ta layer is used mostly for film adhesion and for its reversed sign of ISHE coefficient with respect to Pt.  A sheet resistance of about 44-49 $\Omega /$square was measured for this trilayer. The stack is patterned into $w$ x $l$ rectangles by laser lithography and ion bean etching. Then planar antennas of the different shapes are fabricated using laser lithography and lift-off of a Ti 10/Au 60 nm layer. This second lithography is aligned onto the first one so as to ensure electrical contact between the spintronic trilayer and the metal arms of the antennas. For comparison, STE of identical size were fabricated without any antennas.

\begin{figure}
    \centering
    \includegraphics[width=1\linewidth]{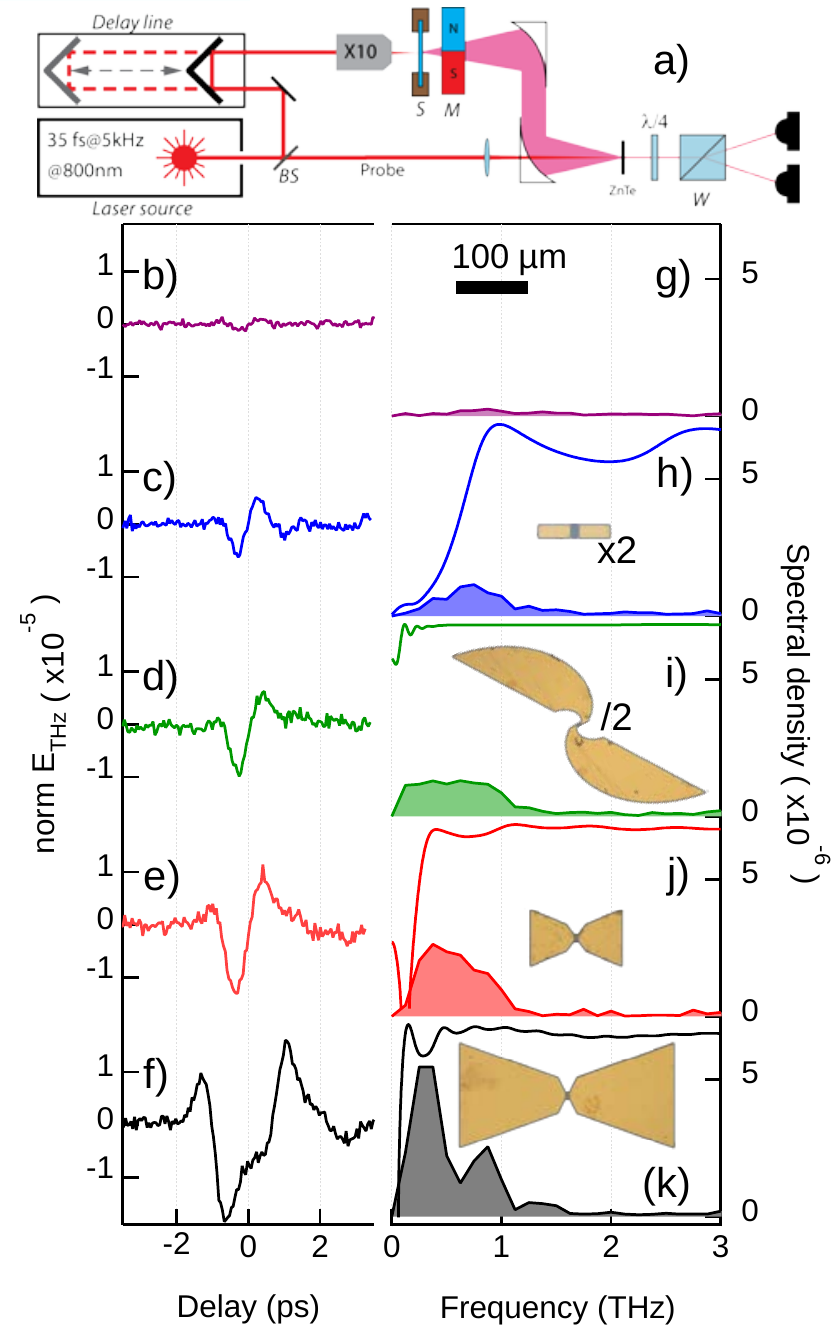}
    \caption{THz signals measured for spintronic TeraHertz emitters equipped with antennas of different shape. (a) Sketch of the pump-probe experimental set-up. (b) THz timetrace measured for a single STE of size 10 µm x 7 µm. (c-f) same for STE equipped with dipole, spiral, compact bow-tie and large bow-tie antennas, respectively. (g-k) Fourier transform of the time trace of panels (b-f) [Solid lines, hidden right scale normalized to 1: Emission coefficient simulated for the different designs, same as Fig.~2(c)]. The zero frequency component of the Fourier transform of each measured waveform has been set to zero. Optical microscope picture of the different antennas are shown as insets with the corresponding scale factors.}
    \label{fig3}
\end{figure}

The THz emission for the different designs has been characterized using a pump-probe technique. A DC magnetic field of $\sim$350mT was permanently applied along the sample plane so as to orient the magnetization of the cobalt layer in-plane, perpendicular to the antenna arms ($y$ in Fig. \ref{Fig1}) Femtosecond pump pulses (800nm@35fs), delivered by an amplified 10kHz laser system, are focused on the single selected device using a X10 objective lens. The spot size of ~30~µm ensures the homogeneous pump fluence of 2.8 mJ/cm$^2$ over the emitter surface, maximizing the spin current generation. The generated sub-ps electromagnetic pulses are collected from the substrate side and focused onto a ZnTe crystal via two parabolic mirrors. An Electro-Optical Sampling (EOS) technique is used to retrieve the electric field profile of the THz pulse ($E_{THz}$). The transient birefrigence of ZnTe is proportional to the instantaneous electric field of the propagating THz pulse. Thus, by varying the delay between the THz and co-propagating probe pulses the THz electric field profile can be retrieved by analyzing the probe polarization [Fig.\ref{fig3}(a)]. The $E_{THz}$ signal displayed further corresponds to the normalized EOS signal with respect to the sum of the static values of the two arms of the polarization bridge.   
Fig.~\ref{fig3}(b) shows the evolution of normalized EOS signal ($E_{THz}$) as a function of pump-probe delay, when the pump is focused on a single 10~x~7~µm$^2$ emitter. The timetrace displays a weak bipolar structure with a peak to peak amplitude of the order of $\sim$1.7x10$^{-6}$ and a a peak to peak timeshift of about 0.5ps. Fig.~\ref{fig3}(c) shows the time-trace measured when the pump is focussed onto a dipolar antenna.  Interestingly, the signal amplitude of $\sim$1.1x10$^{-5}$ is 7 times more intense than the one measured for the single emitter. This illustrates the basic effect of the antenna: for the same pump energy / emitter size, and therefore the same local generated current, the antenna increases significantly the effective length for radiation in free space, and therefore the amplitude of emission.\cite{Kraus_book} Fig.~\ref{fig3}(d,e,f) show the signals measured when the same pump is focused onto spiral, compact, and large bow-tie antennas, respectively. Compared to the dipole antenna, the spiral antenna shows an increased amplitude ($\sim$1.5x10$^{-5}$ peak to peak) together with a slightly modified shape (more unipolar). In a similar way, the compact bow tie antenna shows an even larger amplitude increase up to $\sim$2.1x10$^{-5}$ peak to peak. Finally the time-trace for the large bow-tie antenna displays the strongest amplitude ($\sim$3.4x10$^{-5}$ peak to peak), with a multi-pulse behaviour and a longer overall timescale ($>$ 2~ps). We checked that the time-traces flip sign when the magnetic field is reversed, in accordance with the symmetry expected for a Spin Hall effect, which confirms the STE origin of the signal.

The histograms in Fig.\ref{fig3}(g-k) display the Fourier transforms of the signals measured. Qualitatively, one recognizes the narrow-band behaviour of the dipole antenna, with an emission peak centered at around 0.75~THz and a 60 percents pass band of the order of 0.5~THz. The three other designs display about 50\%  enhanced amplitude at this frequency, and most spectacularly very large enhancement at lower frequencies (enhancement factors of about 2, 4.5 and 9 around 0.3~THz for the spiral, compact bow-tie and large bow-tie antennas, respectively, with respect to the dipole antenna).
Lines in Fig.\ref{fig3}(g-k) show the emission fraction determined by electromagnetic simulations. Overall, the measured low frequency behaviours are in agreement, with relatively large amplitude below 0.8 THz observed for both compact and large bow-tie antennas, and more notably, a very flat spectrum below 1 THz for the  spiral antenna. The dipole antenna shows a clear decay of amplitude as predicted by the simulation.  On the contrary, high frequency components appear much weaker in all measurements than in the corresponding simulations, with an extinction observed above 1.5 THz. We attribute this difference at high frequency to the combination of several factors. The first factor is directivity: the THz parabolic mirrors used have an aperture angle of 30 degrees, whereas the simulated emission corresponds to an emission over the whole space. Simulated radiation diagrams indicate an emission on the substrate side with a significant amplitude out of the substrate normal. This effect is particularly strong at high frequencies and for large antennas (see e.g. the 2.5 THz radiation diagram of the spiral antenna in Fig.\ref{Fig2}(c), where the maximum emission is predicted at about 45° from the inward substrate normal). The second factor is electromagnetic loss associated with metal arms and substrate,  which has been neglected in simulations. Such loss increase generally as function of frequency.\cite{Tydex} The third factor is the finite thickness of the substrate, likely to lead to reflection, refraction and/or wave trapping effects,\cite{Brueck2000} which further reduce the fraction of amplitude reaching the detector, particularly at high frequencies. These three factors, combined with additional limitations of the set-up (0.5~mm thickness of the ZnTe crystal) lead to a limited effective bandwidth in our experiment, which also explains the shift of maximum emission of the dipole antenna measured at about 0.75~THz instead of the 1~THz target. 

\begin{figure}
    \centering
    \includegraphics[width=1\linewidth]{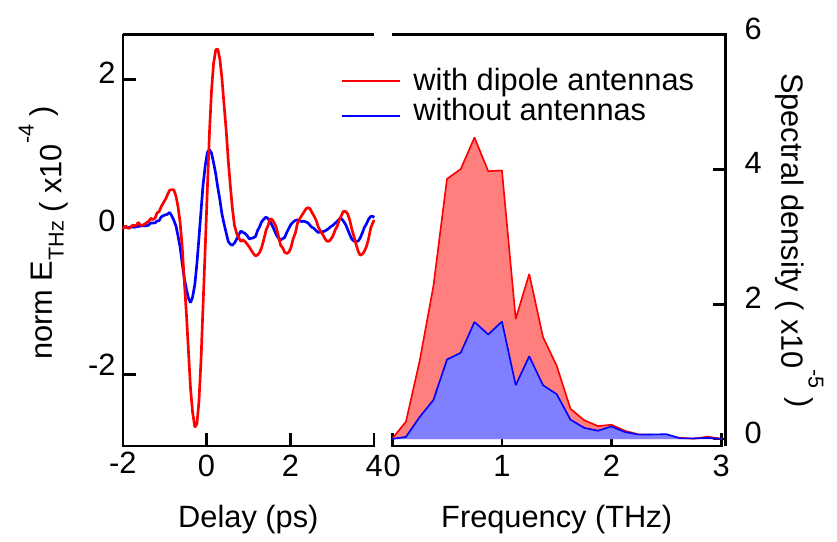}
    \caption{(left) THz time-trace signals measured for arrays of spintronic TeraHertz emitters with or without dipole antennas (red and blue line, respectively) . (right) Fourier transforms of the timetraces. Pump diameter is about 256 µm, with a fluence similar to that for single emitter measurements.}
    \label{fig4}
\end{figure}

Finally, we investigated the emission of arrays of antennas. For this purpose, we increased the pump diameter to 256~µm and increase the pump power accordingly to keep a comparable fluence. The red curve in Figure~\ref{fig4} shows the timetrace measured for an array of dipole antennas (same geometry as individual dipole antenna), with periods of 67~µm and 20~µm in the directions parallel and perpendicular to antenna arms, respectively. For reference, the blue line shows the time trace measured for a similar array in the absence of antenna arms. One recognizes a clear increase of the peak to peak amplitude , from 2x10$^{-4}$ for the array of bare emitters to 5.1x10$^{-4}$ for the array of emitters equipped with dipole antennas. The corresponding enhancement factor amounts to 2.6. This enhancement factor is smaller than the one measured in the case of the isolated dipole antenna (7), which we attribute to an array effect, affecting the directivity of the emission to a different extent depending on the effective length of the emitters.\cite{Kraus_book} Despite this difference of magnitude, the measured enhancement confirms the increase of the far field emission of a small-size spintronic emitter upon insertion in a suitable planar antenna geometry.

To conclude, we demonstrated the positive influence of planar antenna arms onto the far field THz emission of small size THz spintronic emitters and validate a methodology to design such antennas. This work opens the way for further developments of THz spintronics resorting to electromagnetic wave engineering, including a more global optimization, the use of advanced planar circuits and/or non planar designs.

The authors thank the STnano platform for assistance with nanofabrication and Martin Bowen and Jon Gorchon for stimulating discussions. This work of the Interdisciplinary Thematic Institute QMat, as part of the ITI 2021-2028 program of the University of Strasbourg, CNRS and Inserm, was supported by IdEx Unistra (ANR 10 IDEX 0002), and by SFRI STRAT’US project (ANR 20 SFRI 0012) and EUR QMAT ANR-17-EURE-0024 under the framework of the French Investments for the Future Program. This work was also supported by the R\'{e}gion Grand Est and European Union FEDER program through the NanoTerahertz project and by the France 2030 PEPR Spin program through project ANR-22-EXSP-0003 TOAST.

\bibliographystyle{aipnum4-1}
\bibliography{Spintronic_TeraHertz_emitters_with_antennas}

\newpage

\end{document}